\documentclass[a4paper]{article}

\usepackage{atmohead2013}
\usepackage[english]{babel}
\usepackage{verbatim}
\usepackage{graphicx}
\usepackage{amssymb}
\usepackage{multirow}
\usepackage{tabularx, colortbl}

\title{Status and motivation of Raman LIDARs development for the
  CTA Observatory} 

\shorttitle{CTA Raman LIDARs}

\authors{
M.~Doro$^{1,2,3}$,
M.~Gaug$^{2,3}$,
J.~Pallotta$^{4}$,
G. Vasileiadis$^{5}$,
O.~Blanch$^{6}$,
F. Chouza$^{4}$,
R. D'Elia$^{4}$,
A. Etchegoyen$^{7}$,
Ll. Font$^{2,3}$,
D. Garrido$^{2,3}$, 
F.~Gonzales$^{4}$,
A. L\'opez-Oramas$^{6}$, 
M. Mart\'inez$^{6}$, 
L. Otero$^{4}$
E.~Quel$^{4}$,
P. Ristori$^{4}$,
for the CTA consortium 
}

\afiliations{
$^1$ Univ. and INFN Padova, via Marzolo 8, 35131 Padova (Italy),
  $^2$ F{\'i}sica de les Radiacions, Departament de F{\'i}sica, 
  Universitat Aut{\`o}noma de Barcelona, 08193 Bellaterra, Spain,  
$^3$  CERES, Universitat Aut{\`o}noma de Barcelona-IEEC, 08193 Bellaterra,
  Spain, 
$^4$ CEILAP (CITEDEF-CONICET), UMI-IFAECI-CNRS (3351) - Buenos Aires,
Argentina, 
$^5$ LUPM. Montepellier, France, 
$^6$ IFAE, 08193 Bellaterra, Spain,
$^7$ ITeDA (CNEA CONICET - UNSAM) - Buenos Aires, Argentina.
}

\email{michele.doro@pd.infn.it}

\abstract{The Cherenkov Telescope Array (CTA) is the next generation
  of Imaging Atmospheric Cherenkov Telescopes. It would reach
  unprecedented sensitivity and energy resolution in very-high-energy
  gamma-ray astronomy. In order to reach these goals, the systematic
  uncertainties derived from the varying atmospheric conditions 
  shall be reduced to the minimum. Different instruments may help to
  account for these uncertainties. 
  Several groups in the CTA consortium are currently building 
  Raman LIDARs to be installed at the CTA sites. Raman
  LIDARs are devices composed of a powerful laser that shoots into the
  atmosphere, a collector that gathers the backscattered light from
  molecules and aerosols, a photosensor, an
  optical module that spectrally select wavelengths of interest, and a
  read-out system. Raman LIDARs can reduce 
  the systematic uncertainties in the 
  reconstruction of the gamma-ray energies down to 5\%
  level. 

  All Raman LIDARs subject of this work, have design features that
  make them different than typical Raman LIDARs used in atmospheric
  science, and are characterized by
  large collecting mirrors ($\sim2~$m$^2$). They have
  multiple elastic and Raman read-out channels (at least 4) and
  custom-made optics design. In this paper, the motivation for Raman
  LIDARs, the design and the status of advance of these technologies
  are described.}  


\begin{document}
\maketitle

\section{Introduction}
Currently in its design stage, the Cherenkov Telescope Array (CTA) is
an advanced facility for ground-based very-high-energy  gamma-ray
astronomy~\cite{bib:actis}. It is an international initiative to
build the next-generation Cherenkov telescope array covering the
energy range from a few tens of GeV to a few hundreds of TeV with an
unprecedented sensitivity. The design of CTA is based on currently
available technologies and builds upon the success of the present
generation of ground-based Cherenkov telescope arrays (H.E.S.S., MAGIC
and VERITAS\,\footnote{\url{www.mpi-hd.mpg.de/hfm/HESS/},\\
  \url{wwwmagic. mppmu.mpg.de}, \\\url{veritas.sao.arizona.edu}}).  

Nowadays, the main contribution to the systematic uncertainties  
of imaging Cherenkov telescopes stems from the uncertainty in the
height- and wavelength-dependent atmospheric transmission for a
given run of data. Atmospheric quality affects the measured Cherenkov
yield in several ways: the air-shower development itself, the loss of 
photons due to scattering and absorption of Cherenkov light out of
the camera field-of-view, resulting in dimmer images and the
scattering of photons into the camera, resulting in blurred images. 
Despite the fact that several supplementary instruments are currently
used to measure the atmospheric transparency, their data are only used
to retain good-quality observation time slots, and only a minor effort
has been made to routinely correct data with atmospheric
information~\cite{bib:nolan1,bib:dorner,bib:reyes}.  
This situation may change with the CTA atmospheric monitoring and
calibration program~\cite{Doro:2013swa}. There are several goals behind this program. The
first is to increase the precision and accuracy in the energy and flux reconstruction throughout the
use of one or more atmospheric instruments. Secondly, a precise and
continuous monitoring of the atmosphere will allow for an increase of
the telescope duty cycle with the extension of
the observation time during hazy atmospheric conditions, which are
normally discarded in the current experiments because of the
uncertainty in the data reconstruction. Finally, a possible ``smart
scheduling'', i.e., an adaptation -- if required -- of the observation
strategy during the night that considers the actual atmospheric
condition can be activated with a precise monitoring program of the
atmospheric conditions.

In this document, we mainly focus on the importance of the use of
Raman LIDARs to retrieve the differential atmospheric absorption at
different wavelengths at the observatory site. We will show why the
Raman LIDAR is expected to be an optimal instruments for CTA. There
are currently 
three groups developing independently Raman LIDARs for CTA, but the
interest around these instruments is increasing and possibly other groups
will try to develop different solutions. In this document, we present
the main design ideas behind these LIDARs and the progress in their
construction. 

\vspace{-3mm}
\section{Why CTA needs Raman LIDARs}

\begin{figure}
\centering
\includegraphics[width=0.8\linewidth]{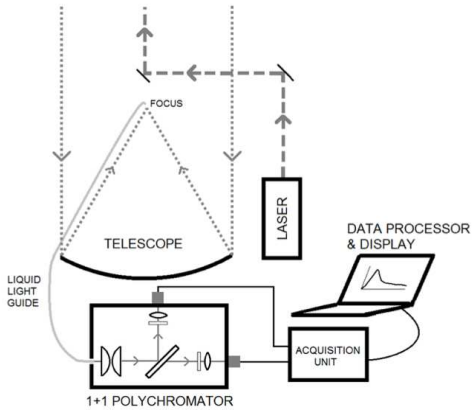}
\vspace{-3mm}
\caption{\label{fig:lidar2}Schematic view of a possible Raman LIDAR
  for the CTA. A laser is pointed towards the atmosphere, and the
  backscattered light collected by a telescope. At the focal plane, a
  light guide transports the light to a polychromator unit which is 
  controlled and readout by an acquisition system and a data processor
  unit.}
\end{figure}

LIDAR is an acronym for {\underline LI}ght {\underline D}etection
{\underline A}nd {\underline R}anging. 
The methodology of the LIDAR technique requires the
transmission of a laser-generated light-pulse into the atmosphere (see
Fig.~\ref{fig:lidar2}). The
amount of laser-light backscattered into the field of view of an
optical receiver on the ground, is then recorded and analyzed. 
LIDARs have proven to be a powerful tool for environmental
studies. Successful characterization of the atmosphere has been made at night using these
systems~\cite{bib:inaba, bib:ansmann, bib:behrendt},
and in other fields of astronomy, the LIDAR technique has proven to be
useful for the determination of the atmospheric extinction of starlight~\cite{bib:zimmer}.
Of the various kinds of LIDARs, the so-called elastic one make only use of
the elastically backscattered light from atmospheric molecules and particles, while
the Raman LIDARs make also use of the backscattered light from
roto-vibrational excitation of the sole atmospheric molecules and not
the aerosols. 
Elastic LIDARs are the simplest class of LIDAR, but their backscatter
power return depends on two unknown physical quantities (the total
optical extinction and backscatter coefficients) which need to be
inferred from a single measurement. As a result various assumptions
need to be made, or boundary calibrations introduced, limiting the precision of 
the height-dependent atmospheric extinction to always worse than 20\%. The introduction of 
additional elastic channels and/or Raman (inelastic-scattering)
channels allows for
simultaneous and independent measurement of the extinction and
backscatter coefficients with no need for \emph{a priori}
assumptions~\cite{bib:ansmann}. Raman LIDARs yield a precision of the
atmospheric extinction at the 5\% level.

The LIDAR return signal can be described by the LIDAR
  equation:
\begin{equation}\label{eq:lidar}
P(R,\lambda_{rec})=K\;\frac{G(R)}{R^2}\;\beta(R,\lambda_{em})\;T^\uparrow(R,\lambda_{em})\;T^\downarrow(R,\lambda_{rec})\quad,
\end{equation}
which  contains a system factor $K$ (emitted
power, pulse duration, collection area of the telescope),  a
geometrical overlap factor (overlap of the telescope field-of-view with
the laser light cone) $G(R)$,  the molecular and aerosol backscatter
coefficient $\beta(R,\lambda_{em})$ and the transmission terms
$T^{\uparrow}(R,\lambda_{em})$ and
$T^{\downarrow}(R,\lambda_{rec})$. $R$ is the atmospheric range,
i.e. the distance from the LIDAR optical receiver, and
$\lambda_{em,rec}$ are the emitted and received wavelengths.

Using the elastic  and Raman-scattered profiles, the
atmospheric aerosol extinction coefficients $\alpha^{m,p}$ ($m$ stands
for molecules and $p$ stands for particles or aerosol) can be derived
with good precision. In this case, the inversion equation has only the
so-called \AA
ngstr\"om index as free parameter (if only one elastic-Raman
wavelength pair is used) and over- or underestimating 
the \AA ngstr\"om index by 0.5 leads to only 5\% relative error in the
extinction factor. Hence, apart from statistical uncertainties (which
can be minimized by averaging many LIDAR return signals), results are
typically precise to about 5-10\% {\bf in each altitude bin}, and
probably even better in clear free tropospheres with only one aerosol layer.
The uncertainty generally grows with increasing optical depth of the
layer. By adding a {\bf second Raman line}, e.g. the $N_2$ line at
607~nm, the last free \AA ngstr\"om parameter becomes fixed, and
precision of {\bf better than 5\%} can be achieved for the aerosol
extinction coefficients. The molecular extinction part needs to be
plugged in by hand using a convenient model. However, since the
molecular densities change very little, and on large time scales,  
this can be achieved by standard tools. Precision of typically better than 2\%
are rather easy to achieve. \\

Although IACTs are normally placed at astronomical sites,
characterized by extremely good atmospheric conditions, the local
atmosphere is potentially influenced by phenomena occurring at tens to
thousands of kilometers away, and thus should be continuously
monitored. 
While the molecular content of the atmosphere varies very slowly at a
given location during the year, and slowly from place to place,
aerosol concentrations can vary on time-scales of minutes and travel
large, inter-continental, distances. Most of them are concentrated
within the first 3~km of the troposphere, with the
free troposphere above being orders of magnitude cleaner. 
Aerosol sizes reach from molecular dimensions to millimeters, and the particles
remain in the troposphere from 10 days to 3 weeks. The sizes are
strongly dependent on relative humidity. 
Different types of aerosol show characteristic size distributions, and
an astronomical site 
will always show a mixture of types, with one possibly dominant type at a
given time and/or altitude.
Light scattering and absorption by aerosols needs to be described by
Mie theory or further developments of it, including  non-sphericity of
the scatterer. Aerosols generally have larger refraction indexes than that of water,  and typically show also a small imaginary
part. Contrary to the typical $\lambda^{-4}$ wavelength dependency of
Rayleigh-scattering molecules, aerosols show power-law indexes (the
so-called \textit{\AA ngstr\"om} coefficients) from  0 to 1.5, i.e. a
much weaker dependency on wavelength. 

In order to estimate the effect of different atmospheric conditions on
the image analysis of IACTs, we have simulated different molecular and
aerosol profiles for the MAGIC system, consisting of two
telescopes. The results were presented in a master thesis and elsewhere in this
conference~\cite{bib:garrido_phd,bib:garrido_icrc}. Several aerosol
scenarios were simulated: $i)$ enhancements of the ground layer from a
quasi aerosol-free case up to  a thick layer which reduces optical 
transmission by 70\%, $ii)$ a cloud layer at the altitudes of 6~km, 
10~km (cirrus)  and 14~km~(volcano debris)~a.s.l., and $iii)$ a 6~km
cloud layer with varying aerosol densities. The main results can be
summarized in three points: 

\begin{enumerate}
\item Using correct MC, energy and flux reconstruction is correct, at
  only the expense of a larger energy threshold, which can be
  explained by the fact that with hazy atmospheres fewer photons reach
  the ground;
\item In the case that the aerosol overdensity or cloud is below the
  electromagnetic shower, a simple correction method can be used to
  restore correct energy and flux reconstruction with the simple use
  of standard Monte Carlo;
\item When the clouds or aerosol layer is at the shower development
  region or above, {\bf the total extinction is no longer an useful parameter}
\end{enumerate}

\begin{figure}[h!t]
\centering
\includegraphics[width=0.98\linewidth]{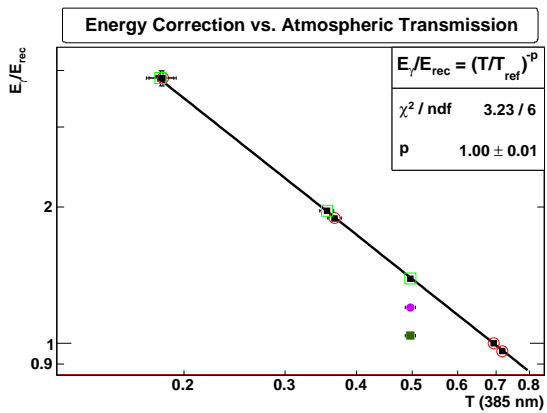} 
\vspace{-3mm}
\caption{\label{fig:ecor}Energy reconstruction as a function of the
  total transmission at 385 nm. The linear dependence is valid only in
  the case of low aerosol overdensities. $E_{\rm rec}$ is the reconstructed
  energy from an initial Monte Carlo event with energy $E_\gamma$. $T_{\rm ref}$ is a scale transmission at 385~nm and
  $-p$ the chi-square optimized index, $T$ is the transmission at
  385~nm for each of the points in the plot. See text for additional details. Taken
  from~\cite{bib:garrido_icrc}.} 
\end{figure}

The last two points are shown in Fig.~\ref{fig:ecor} where the energy
correction is shown as a function of the total transmittance at
385~nm. One can see that in the case that the aerosol overdensity is located
close to the ground, i.e. below the electromagnetic shower
development, then the energy reconstruction is precisely obtained from
the total transmission, while if the overdensity is above, then
correlation is broken and {\bf height-resolving instruments are
  required}. This is the main motivation for the need of a Raman LIDAR
instrument for CTA. 
We believe that the main findings of this study should also
be valid for CTA, at least in the energy range from 50 GeV to 50 TeV. Previous studies have been made
\cite{bib:bernloher,bib:nolan1,bib:dorner} for H.E.S.S. and
for the MAGIC mono system, however only for an increase of low-altitude
aerosol densities, and in \cite{bib:nolan2} for a reference
configuration of CTA, claiming a change in the spectral power-law
index of  gamma-ray fluxes, when atmospheric aerosol layers are
present. In our work, we found that different atmospheres affect the
energy threshold, the energy resolution and the 
energy bias, that propagate into the computation of a target flux and
spectral reconstruction. See \cite{bib:garrido_phd,bib:garrido_icrc} for further details. 

Atmospheric properties can be derived, to a certain extent, also directly
from IACT data.  Several studies have been made by the H.E.S.S. and
MAGIC collaborations to estimate the integral atmospheric
transmission, using trigger rates, muon rates, combinations of both~\cite{bib:reyes},
or the anode currents of the photomultipliers and/or pedestal RMS. Up
to now, these parameters have been used only to discard data taken
under non-optimal conditions, but work is ongoing to use this
information to also correct data. However, as stated above, the
use of integral transmission parameters is only valid in some of the
possible atmospheric scenarios. In addition, data obtained directly
from the telescopes suffer from the fact that instrumental properties
(e.g. mirror reflectivity, PMT aging, etc) could influence these
data, and therefore the separation between purely atmospheric effects
could be affected. Anyhow, we have investigated the possibilities of
using remote sensing devices such as the LIDAR.  
The experience of MAGIC with an elastic LIDAR system (i.e. analyzing
only one backscatter wavelength, and no Raman lines),   has shown that
simplified reconstruction algorithms can be used to achieve
good precision of the aerosol extinction coefficients, at least within  
the range of uncertainties inherent to an elastic LIDAR \cite{bib:fruck}. 
An analog conclusion was achieved with the H.E.S.S. LIDAR: a stable
analysis algorithm was found, limited by the 30\% uncertainties of the
time and range dependent LIDAR ratio.

\vspace{-3mm}
\section{Raman LIDARs prototypes}
Several institutes in CTA are currently designing Raman LIDAR systems:
the Institut de F\`isica d'Altes Energies (IFAE) and the Universitat
Aut\`onoma de Barcelona (UAB), located in Barcelona (Spain), the LUPM
(Laboratoire Univers et Particules de Montpellier) in Montpellier
(France) and the CEILAP (Centro de Investigaciones Laser y sus
Aplicaciones) group in Villa Martelli (Argentina). The different
groups are designing independently the LIDAR systems with different
mechanical, optical and steering solutions. In addition, two other
groups have shown interest in discussing new designs for Raman LIDARs,
one in Adelaide (Australia) and one from INFN (Italy). We anticipate
that these designs are not necessarily in competition, and that the
fact that CTA will have two different sites, one in the Northern
and one in the Southern hemisphere, and that CTA could be
operated also with different sub-arrays pointing at different
directions in the sky, could possibly require the use of several Raman
LIDARs at the same site. However, the overall strategy is not defined
yet. Some information about the different LIDAR is collected in
Table~\ref{tab:lidars}. 

\begin{table*}[h!t]
\centering
\small{%
\begin{tabular}{lccc}
\hline
\rowcolor[gray]{.8} 
& {\bf CEILAP} & {\bf IFAE/UAB} & {\bf LUPM} \\
\hline
Housing & Custom-made container & CLUE container & CLUE container \\
Design & Multi-angle & Co-axial & Co-axial \\
Mirror diameter [cm] & 6x40 & 1x180 & 1x180 \\
Mirror f/D & 2.5 & 1 & 1 \\
\hline
Elastic 355 nm       & X & X & X \\
Raman 387 nm (N$_2$) & X & X & X \\
Raman 408 nm (H$_2$O)& X &   &   \\
Elastic 532 nm       & X & X & X \\
Raman 607 nm (N$_2$) & X & X & X \\
Elastic 1064 nm      & X &   &   \\
\hline
Laser & Continuum Inlite-II & Quantel Brilliant & Quantel CFR400\\
Max. Laser power [mJ/p] & 125 @ 1064 nm & 100 @ 355, 532 nm & 90\\
Pulse duration [ns] & $7-9$ & 5 & 7 \\
Beam diameter [mm] & 6 & 6 & 7 \\
Beam divergence [mrad] & $<0.75$& 0.5 & 3.5 \\
Pulse frequency [Hz] & 50 & 20 & 20 \\
\hline
Light transport & Optical fiber & Liquid lightguide & Liquid
lightguide \\
                & Edmund UV/VIS series  & Lumatec Series 300 &
Spectral Labs.\\
Fiber size [mm] & 1 & 8 & 8 \\
Fiber N.A. [mm] & 0.22 & 0.64 & 0.6 \\
\hline
Photon detector & PMT & PMT/HPD & PMT \\
& Hamamatsu H10721-110$^*$ & Hamamatsu R11920-100$^*$ & Hamamatsu R329P\\ 
Cathode diameter [inch] & 1 & 1.5 & 2 \\
\hline
Readout system  & LICEL & LICEL & LICEL \\
\hline
\end{tabular}
}
\vspace{-3mm}
\caption{\label{tab:lidars}Data collection of the Raman LIDARs for
  CTA. $^*$Subject to change.}
\end{table*}

\subsection{CEILAP design\\ \small{Contact: P.Ristori (pablo.ristori@gmail.com)}}
\begin{figure}[h!t]
\centering
\includegraphics[width=0.90\linewidth]{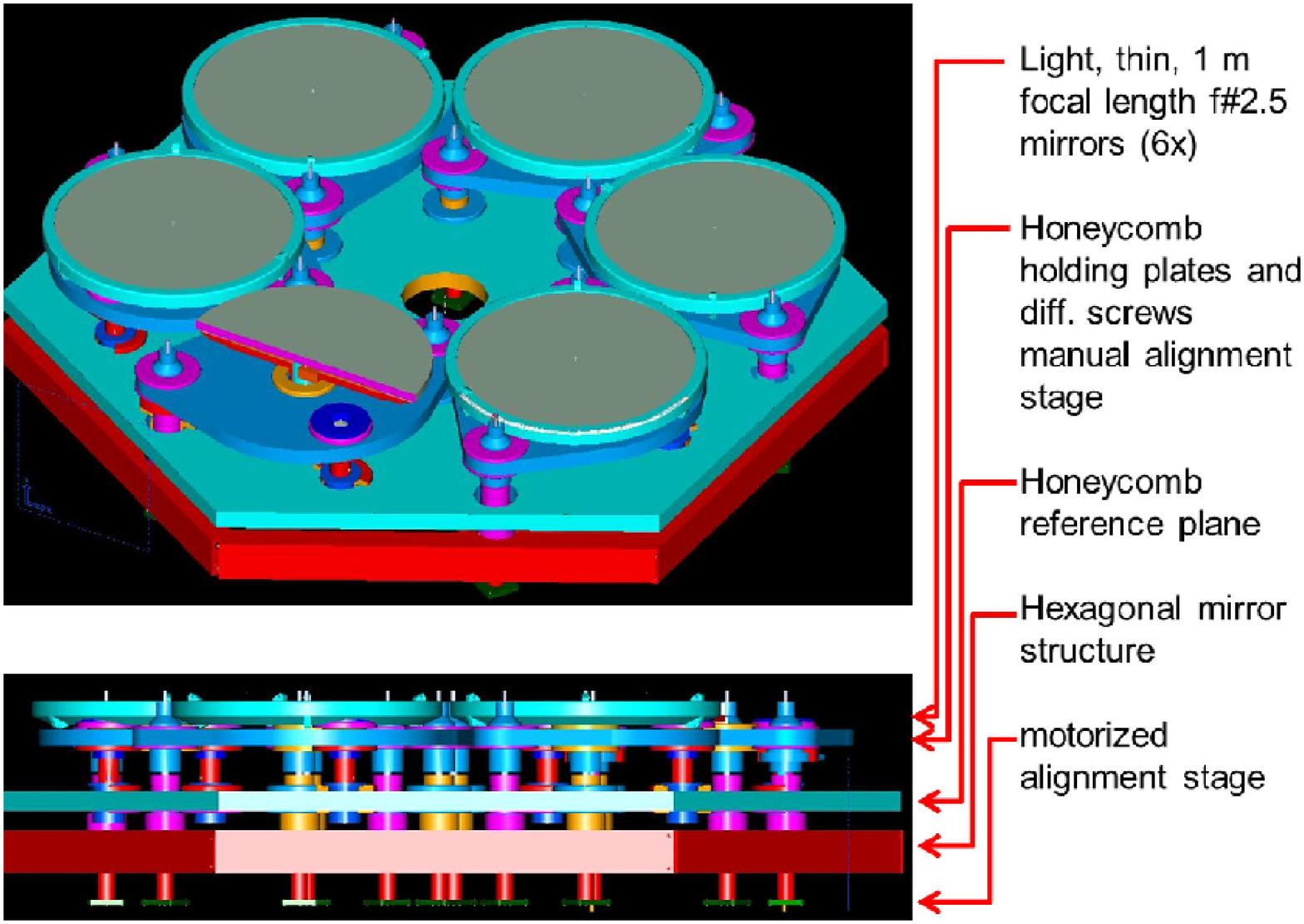} 
\includegraphics[width=0.90\linewidth]{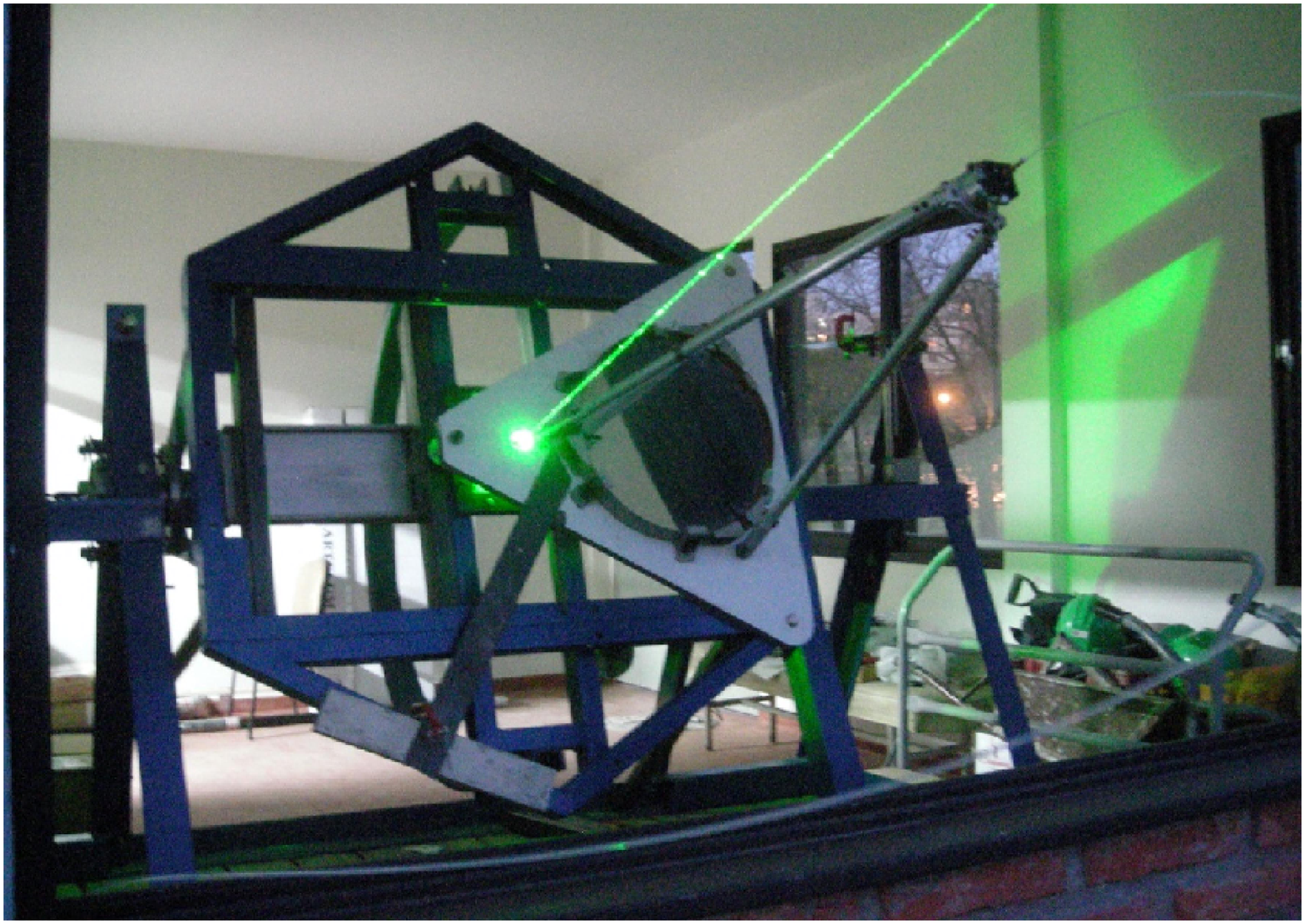} 
\vspace{-3mm}
\caption{\label{fig:ceilap} (top) Multi-mirror setup of the LIDAR system from
  the top and lateral view showing the different stages of the
  system. (bottom) The first sector of the LIDAR mounted and tested.}
\end{figure}

The CEILAP group is developing a fully custom-made multi-wavelength
scanning Raman LIDAR in the frame of the Argentinean CTA
collaboration. This LIDAR emits 
laser pulses of $7-9$ ns duration at 355, 532 and 1064 nm and at 50 Hz with
nominal energy of 125 mJ at 1064 nm and records these wavelengths and
the two nitrogen Raman shifted wavelengths from 532 nm (607 nm) and 355 nm
(387 nm). A sixth wavelength at 408 nm is also used to detect the
water vapor Raman return. The system is peculiar because it uses six
commercial 40~cm diameter parabolic mirrors with $f/2.5$ that
can be used as six independent Newtonian telescopes. This permits to
avoid dealing with typical deformation, 
aberration issues and higher costs that are usually connected with
larger mirrors. This solution also permits to possibly 
extract any mirror for recoating or exchanged keeping the other
five mirrors operating. Furthermore mirror construction and coating can be done by
standard methods. The main drawback of the chosen solution is that
each of the six telescopes must be aligned properly to attain the maximum
system efficiency. The additional alignment procedure
has been solved by an automatic mirror alignment to follow the line of
sight of the observation during the acquisition period. For this, each
mirror is equipped with actuators on their backsides. The system was
designed to operate in hard environmental conditions. A image of the
LIDAR is shown in Fig.~\ref{fig:ceilap}.

The multi-mirror telescope unit was designed to provide a maximum
stability to the system with a minimum weight. While honeycomb was
used for the multi-mirror reference plane, carbon fiber tubes were
used to place the optical fiber at the mirror focal plane. Nylon
pieces were synterized at the end of the carbon tubes to provide
better fixation. The laser model is Inlite II-50 from Continuum. It
has a hardened design and compact size for reliable operation in
industrial environments. The flash-lamp can be easily removed for
future maintenance. It has a cast aluminum resonator structure which ensures
long-term thermal and mechanical stability. The optical fibers are
standard 1~mm size. The optical module unit is totally custom-designed
and custom-made, with the main structure being built with 3d printing
machines. For the PMTs, bialkali (UBA) or superbialkali (SBA) modules
from Hamamatsu are foreseen. Possible choices will be H10721-220 (UBA)
o H10721-110 (SBA). During 2012 a new shelter-dome to host the LIDAR was
acquired. A standard 20 ft container was cut and equipped with
hydraulic pistons. 
The whole LIDAR is remotely
controlled using a wi-fi link from the control PC to the LIDAR shelter
creating a local LIDAR network under the TCP/IP protocol. Both
acquisition and shelter controls will be operated remotely by the
shifter. 
This control system is fully functional. 
The operation scheme
is presented in \cite{Pallotta:2013atmohead}. 


The first sector of the LIDAR is currently mounted and under
tests together with the laser and the alignment system. The final telescope steering system is under development as
well as the motorized azimuth-zenithal mechanism in acquisition
process~\cite{Pallotta:2013atmohead}. The next steps are designing the spectrometric box and
building the electronic control in its final version. For the optical
module, photomultipliers from Hamamatsu H10721-110 are foreseen for
all channels. 

\newpage
\subsection{IFAE/UAB design\\ \small{Contact: M.Gaug (markus.gaug@uab.cat)}}
\begin{figure}[h!t]
\centering
\includegraphics[width=0.90\linewidth]{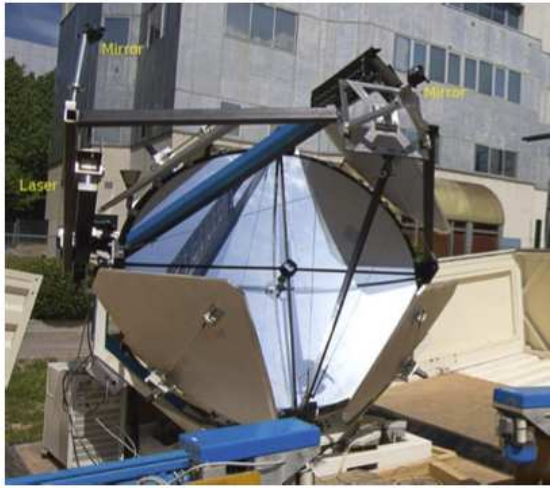} 
\vspace{-3mm}
\caption{\label{fig:bcn}The Barcelona IFAE/UAB Raman LIDAR. A similar
  LIDAR is being assembled by LUPM.}
\end{figure}

Fig.~\ref{fig:bcn} shows an image of the Barcelona Raman LIDAR. The
telescope structure and housing container has been inherited from the
former CLUE experiment  and is currently being refurbished for its
adaptation to a LIDAR system~\cite{bib:lopez}. It is composed of a
steel structure, over which a primary mirror is fixed. The mirror is a
solid glass parabolic dish of 1.8~m diameter with an $f/1$ and thickness of 6 mm obtained with
hot slumping technique developed at CERN. The point spread function of
the mirror is about 6~mm while its reflectivity has degraded compared
to the CLUE times and is currently about 60\%. For this reason the mirror will be
refurbished in the future. The telescope was equipped with
a Nd:YAG laser from Quantel Brilliant emitting at the primary
wavelength of 1064 nm with the first two harmonics available at 532
and 355 nm. The energy per pulse is 60 mJ, the pulse repetition
frequency is 20 Hz and the pulse duration 5 ns, with a beam waist of 6
mm and divergence of 0.5 mrad. The laser is mounted at the side of the
telescope, as shown in Fig.~\ref{fig:bcn} and two small guiding
mirrors mounted in front of the laser, guide the laser light to the
optical axis of the telescope. The LIDAR is therefore called
mono-axial. At the focal plane, a liquid
light-guide from Lumatec Series 300 collects the light of the
mirror. It has an 8 mm diameter section, a numerical aperture of 0.59
and transitivity of more than 0.7 in UVs for the whole length.  The guide was tested and
its performance are quite stable over the working temperature
range. The main problem is that it cannot transport the 1064~nm line
from the laser, which needs to be masked. In addition, the numerical
aperture is relatively large due to the large size of the fiber.
The liquid light-guide
transports the light to an optical unit called a polychromator which
collimates the beam and transports the light to four photomultipliers
for the four wavelengths of interest (the two elastic at 355 and 532
nm and two N2 Raman lines at 387 and 607 nm)~\cite{bib:dadeppo}. The
polychromator will be mounted in the back of the telescope structure
and will move together with the telescope. The current baseline
design foresees PMTs from Hamamatsu. We will use for some channels
Hamamatsu R11920-100 of 1.5 inch size (those used for CTA LST
cameras), and possibly Hamamatsu H10425-01 for the 607 nm line,
because it has a larger efficiency. However, the possible use of
custom Hybrid Photon Detectors (HPD) is still under consideration. The
needed cathode size either for PMT or HPD is at least 20~mm
diameter. Dichroic mirrors and collimating lenses are already acquired
and the assembly is currently under tests now. The acquisition unit is
based on commercial LICEL$^{\rm TM}$ modules.  The telescope steering
is tested and fully functional. The next step is the assembly of the
polychromator unit and the test of the full LIDAR system. Primary
tests will be done in Barcelona and fields test will be done in La
Palma at the MAGIC telescope site.

\subsection{LUPM design\\
 \small{Contact: G. Vasileiadis (george.vasileiadis@lpta.in2p3.fr)}}
The LUPM design is similar to that of the IFAE/UAB Raman
LIDAR. It is also based on the use of the CLUE container and telescope
structure. However, different solutions for the laser, the telescope
and container steering, readout wavelengths, etc. were chosen. In
particular, due to age and relatively difficulty to obtain parts, LUPM
opted for a completely rebuilt of all mechanical parts of the
container. New motors were purchased and installed both for the
container doors operations and the telescope movement. A completely
new automated system, based on the Rockwell automation PowerFlex
series, will take care of the drivers guidance and mechanical operation
of the LIDAR. A dedicated Ethernet based software environment will
assure the user operation. The communication with the rest of the CTA
DAQ system will be done over the OPC/UA protocol. 

Due to the harsh conditions expected for the southern hemisphere CTA
observatory where some of the LIDARs will be installed, LUPM opted for an
industrial/military-type of laser. Their choice was a CFR400 laser
fabricated by Quantel. The CFR laser is a lamp pumped Nd:YAG laser
featuring a degree of ruggedization not found in typical scientific
lasers. The CFR design has been vibration-tested and each laser is
temperature cycled before shipping. It emits at 1064,532 and 355nm
with a peak energy of 90~mJ and 20~Hz repetition rate. It has a
7~ns pulse duration, beam diameter less than 7~mm and beam divergence
better than 3.5~mrad. The laser will be mounted in one of the telescope
masts, while a beam drive system will assure a coaxial operation of
the LIDAR. Collection at the focal point of the telescope will be
achieved using a liquid fiber, Spectra Labs,  of 8~mm diameter, 5~m long
and numerical aperture of $0.6$ with a transmissivity of more than 0.6 in
the UV. A traditional optical box where the necessary Raman filters
and beam splitters and attenuators will be mounted at the back of the
telescope. In total we will detect four wavelengths, with the same
specs as of the IFAE/UAB design. A set of Hamamatsu photomultipliers
tubes of 2 inches diameter will detect the incoming signal, while data
acquisition will be assured by a LICEL commercial system.

\subsection{Other possible contributions}
The atmospheric calibration is raising larger interest in the CTA
community and two additional groups are currently evaluating the
possibility to contribute with independent designs of Raman LIDARs. The
first group relates to the University of Adelaide~\cite{adelaide}
(contact: I. Reid and A. MacKinnon), whose current work is based on
monitoring of greenhouse gases and climate change. The Atmospheric
Physics Group in collaboration with the Optics and Photonics Group is setting up a new LIDAR facility at
Buckland Park. The aim is to measure atmospheric temperature, wind and
dynamical processes with high spatial and temporal resolution from 10
to 110 km altitude. It is possible that they can use the facility to
build a Raman LIDAR for the needs of CTA. A second group is formed by
the INFN (Italian Institute for Nuclear Physics) sections of Napoli,
Padova and Torino (contact: C. Aramo (aramo@na.infn.it)), with a long-standing experience with the Raman LIDAR and the atmospheric calibration for
the Auger experiments. The groups are joining interests for the
construction of Raman LIDARs for CTA. 

\vspace{-3mm}
\section{Complementary instrumentation}
Apart from the Raman LIDAR, complementary devices for atmospheric
characterization and the understanding of the site climatology can be
used. A first class of devices comprises those which provide at least
some profiling of the atmosphere, such as radio sondes, profiling
microwave or infrared radiometers and differential optical absorption
spectrometers. The operating wavelengths of these devices are very
different from those of the Raman LIDAR, and precise conversion of
their results to the spectral sensitivity window of the CTA is
difficult. However, since aerosols are better visible at larger
wavelengths, profiling devices may be used to determine cloud heights
with high precision and their results may be good seeds for the Raman
LIDAR data inversion algorithm. A next class of complementary devices
contains those which measure integral parameters, such as Sun, Lunar
and stellar photometers, UV-scopes and starguiders. Integral optical
depth measurements have become world-wide standards, organized in
networks ensuring proper (cross-)calibration of all devices.  
Spectacular resolutions of better than 1\% can be obtained during the
day, about 2\% with moon, and 3\% under dark night conditions,  at
wavelength ranges starting from about 400~nm. Extrapolations to the
wavelength range between 300 and 400~nm worsens the resolution
again. The precise results from these devices can serve as important
cross-checks of the integrated differential Raman LIDAR transmission
tables. Finally, all major astronomical observatories operate cloud
detection devices, mainly all-sky cameras and/or take advantage from
national weather radars. All-sky cameras have become standardized
within the CONCAM or the TASCA networks, however important differences
in sensitivity to cirrus clouds are reported. The advantage of these
devices are their big field-of-view which allows to localize clouds
over the entire sky and makes possible online adapted scheduling of
source observations. Relatively cheap cloud sensors based on
pyrometers or thermopiles have been tested by the MAGIC collaboration
and the SITE WP of CTA. The calibration of these devices is however
complex and measurements are easily disturbed by surrounding
installations. Recent analysis can be found in \cite{bib:daniel,bib:hahn}. 

\vspace{-3mm}
\section{Conclusions and outlook}
Monte Carlo simulations showed that while data affected by enhancements of the ground layer
can be scaled rather easily up to high levels of extinction, 
this is not the case anymore for (cirrus) clouds at altitudes from 6
to 12~km a.s.l., which create strong energy-dependent effects on the
scaling factors. Moreover, the images from atmospheric showers are distorted depending on the
location of the shower maximum, which varies even for showers of a
same energy. Very high altitude layers, in turn, produce only 
effects on the very low energy gamma-ray showers.  Depending on the
properties of the chosen site for CTA, still to be decided, it would
probably make sense to create $10-20$ typical atmospheric simulations
within these possibilities and interpolate between them.  
The natural solution is the use of (Raman) LIDARs, which
were described in this contribution. However, the use of
complementary instruments that measure integral or differential (in
altitude) atmospheric parameters is possible and envisaged. Once
retrieved the differential atmospheric transparency, different
strategies may be foreseen to accurately and precisely reconstruct data,
ultimately reducing the reconstructed energy and flux uncertainties.
This can be
achieved by re-calibrating the data themselves, either event-wise  
or bin-wise, or by simulating
adapted atmospheres~\cite{Doro:2013swa}.
In addition, it would be possible to increase the duty cycle of the
telescopes by retrieving those data taken during non-optimal atmospheric
conditions which are normally discarded by standard clean-atmosphere
analysis, especially important during e.g. multi-wavelength
campaigns or target of opportunity observations. 

Three independent groups are currently developing custom-made Raman
LIDARs. In Argentina, CEILAP is developing a novel-design 6-mirror
Raman LIDAR, while in Europe, IFAE/UAB in Spain and LUPM in France are
adapting a telescope from the terminated CLUE experiment as Raman
LIDAR. Status and plans were reported in this proceedings.

\vspace{3mm}
\footnotesize{{\bf Acknowledgment:}{ We gratefully acknowledge support from the agencies and organizations 
listed in this page: \url{http://www.cta-observatory.org/?q=node/22}}}


\begin{thebibliography}{99}
\setlength{\itemsep}{2.0pt plus 4.0pt minus 2.0pt}
\bibitem{bib:actis} M. Actis et al., Exper.Astron. 32 (2011) 193-316.
\bibitem{bib:aleksic} J. Aleksi{\'c} et al.,
  Astrop.Phys. 35 (2012) 435-448
\bibitem{bib:aharonian} F. Aharonian et al., A\&A
  457 (2006) 899-915
\bibitem{bib:nolan1} S. J.Nolan et al., Procs. 30th ICRC (2007) 
\bibitem{bib:dorner} D. Dorner et al. A\&A 493
  (2009) 721-725
\bibitem{bib:reyes} R. De Los Reyes et al, Procs. 33rd ICRC (2013) ID-0610
\bibitem{bib:garrido_icrc} D. Garrido et al., these proceedings.
\bibitem{bib:garrido_phd} D. Garrido. Phd Thesis. Univeristat Aut\'onoma Barcelona (2011)
\bibitem{bib:bernloher} K. Bernl\"ohr, Astrop.Phys. 12 (2000) 255-268
\bibitem{bib:nolan2} S.J. Nolan et al., Astrop.Phys. 34 (2010) 304-313 
\bibitem{bib:inaba} H. Inaba, Springer-Verlag, New York (1976) 143
\bibitem{bib:ansmann} A. Ansmann et al., Appl.Opt. 31 (1992) 7113
\bibitem{bib:behrendt} A. Behrendt et al., Appl.Opt. 41 (2002) 7657 
\bibitem{bib:zimmer} P.C. Zimmer et al., Procs. AAS Meeting, 42 (2010) 401
\bibitem{bib:fruck} C. Fruck et al., Procs. 33rd ICRC (2013) ID-1054.
\bibitem{bib:ristori} P. Ristori et al., Procs. 33rd ICRC (2013) ID-0346.
\bibitem{bib:lopez} A. Lopez-Oramas et al., Procs. 33rd ICRC (2013) ID-0210.
\bibitem{bib:daniel} M. Daniel and G. Vasileiadis, AIP
  Conf. Procs. 1505 (2012) 717-720
\bibitem{bib:hahn} J. Hahn et al., AIP Conf. Procs 1505 (2012) 721-724
\bibitem{Doro:2013swa} M.~Doro {\it et al.} [CTA Coll.],
  Procs. 33rd ICRC (2013) ID-0151
\bibitem{adelaide}http://www.physics.adelaide.edu.au/atmospheric/lidar.html
\bibitem{bib:dadeppo} V. Da Deppo et al., Proc. of SPIE Vol. 8550 85501V-1.
\bibitem{Pallotta:2013atmohead} J. Pallotta et al., 
Procs. of the
  AtmoHEAD Conf., Saclay (2013) 

\end{thebibliography}
\end{document}